\documentclass[aps,prd,twocolumn,showpacs,superscriptaddress,groupedaddress]{revtex4}

\usepackage{graphicx}

\usepackage{dcolumn}
\usepackage{bm}

\usepackage{textcomp} 
\usepackage{amstext} 

\usepackage[utf8]{inputenc}
\usepackage{graphicx}
\usepackage{epsfig}
\usepackage{mathtools, nccmath}
\usepackage{environ}
\usepackage{tabularx}
\usepackage[usenames]{color}

\NewEnviron{myequation}{%
\begin{equation}
\scalebox{0.85}{$\BODY$}
\end{equation}}

\usepackage[colorlinks,filecolor=blue,citecolor=blue,urlcolor=blue]{hyperref}
\usepackage[dvipsnames]{xcolor}
 
\hypersetup{pdfstartview=FitH, linkcolor=Blue,urlcolor=Blue, colorlinks=true}

\RequirePackage{lineno}

\begin{document}

\hspace{5.2in} \mbox{ILD-PHYS-2021-001}

\vspace*{0.1cm}

\title{Measurement of {\boldmath $\sigma(e^+e^- \to HZ) \times {\cal B}r(H \to ZZ^*)$} at the 250 GeV ILC}

\affiliation{P.N. Lebedev Physical Institute of the Russian Academy of Sciences, Moscow 119991, Russia.}

\author{E.~Antonov$^1$, A.~Drutskoy$^1$ \vspace*{0.5cm}}

\begin{abstract}
\vspace*{0.01cm}
We report on studies of the $e^+e^- \to HZ$ process with the subsequent decay of the Higgs boson
$H \to Z Z^\star$, where the $Z Z^\star$ combination is reconstructed in the final states with two jets and two leptons. 
The analysis is performed using Monte Carlo data samples obtained with detailed ILD detector simulation assuming 
the integrated luminosity 2~ab$^{-1}$, the beam polarizations ${\cal{P}}_{e^-e^+} = (-0.8, +0.3)$,
and the center-of-mass energy $\sqrt{s}$ = 250~GeV. The analysis is also repeated for the case of two 0.9~ab$^{-1}$ data samples with polarizations
${\cal{P}}_{e^-e^+} = (\mp0.8, \pm0.3)$. The process is measured in
four decay channels, which correspond to two combinations for the Higgs final states and two decay modes
of the directly produced $Z$ boson, $Z \to q \bar{q}$ and $Z \to \nu \bar{\nu}$.
To obtain the Higgs boson mass distributions, we used
the variables $M(jj\ell\ell)$ and $M_{\Delta} = M(jj\ell\ell) - M(jj) + M(Z_{\rm nom})$, where $M(Z_{\rm nom})$ = 91.2~GeV.
Contributions of the potential background processes are taken into
account based on the available MC event samples.
We propose a model-independent method for obtaining the width of the Higgs boson using the measurement
of the $e^+e^- \to HZ$ process.

\end{abstract}

\smallskip

\pacs{13.38.Dg, 13.66.Fg, 13.66.Jn, 14.80.Bn}

\maketitle

\section{\label{sec:intro}Introduction}

Since the discovery of the Higgs boson by the ATLAS and CMS collaborations~\cite{atlas,cms} in 2012,
the next important task is to measure parameters of this particle with the highest possible accuracy.
In recent years, the mass, couplings, production cross sections, and quantum numbers of the Higgs boson have been measured 
in the LHC experiments with increasing accuracies. However, the width of the Higgs boson
is difficult to measure at LHC in a model-independent approach. Indirect methods will be used to measure 
the Higgs width at LHC after the high-luminosity upgrade,
but even with the data sample of 3~ab$^{-1}$ the uncertainty is expected to be $\sim$20$\,\%$~\cite{ww,pdg}.

The width of the Higgs boson is strictly theoretically defined within the Standard model (SM) for a fixed Higgs mass.
The width value of 4.1~MeV/$c^2$ has been calculated for the mass of 125 GeV/$c^2$~\cite{esw}. This result may
be distorted by beyond the Standard Model (BSM) contributions. Therefore, the model-independent measurement
of the Higgs width provides an important test of SM. 
A high accuracy of the Higgs width measurement can be reached at the future $e^+e^-$ linear collider ILC~\cite{wp,hww}.
A large number of Higgs bosons will be produced at ILC, whereas backgrounds are expected to be relatively small. 
Events in the ILD detector~\cite{ild} at ILC have clean and well-defined signatures and, therefore,
processes of interest can be identified and studied in detail.

We propose to use the process $e^+e^- \to Z H$ with the subsequent decay $H \to Z Z^\star$
to measure the product of the cross section and the decay branching fraction,
which can theoretically be expressed as:

\begin{fleqn}[\parindent]
\begin{equation}\label{eq:sigmaxbr}
\sigma(e^+e^- \to H Z) \times Br( H \to Z Z^\star) = C \cdot {\rm g_Z^4} / \Gamma_H
\end{equation}
\end{fleqn}

\noindent
Here C is a constant which can be calculated theoretically with an uncertainty of less than 1$\,\%$ \cite{hww},
g$_Z$ is the Higgs boson coupling $HZZ$, and $\Gamma_H$ is the Higgs boson width.
Therefore, the measurement of this product can be used to obtain the width of the Higgs boson with a high accuracy, because 
the coupling g$_Z$ is expected to be determined combining ILC and LHC results 
with an uncertainty of about 0.5$\,\%$ \cite{hzzm} using other processes.
In this analysis we assume the data sample of 2~ab$^{-1}$ collected by the ILD experiment
in the $e^+e^-$ collisions at a center-of-mass energy 250 GeV. The accuracy of the $e^+e^- \to Z H$ process
measurement at 250 GeV with the subsequent decay $H \to Z Z^\star$ has also been evaluated using the SiD detector 
simulation~\cite{wp}.

In the studied process the two on-shell $Z$ bosons and one off-shell $Z^\star$ boson have to be reconstructed.
The directly produced $Z$ boson is denoted below as $Z_1$ to separate it from the $Z$ bosons produced in the Higgs decay.
The decays of the $Z_{(1)}^{(\star)}$ bosons to two hadronic jets, two opposite sign leptons ($\ell^\pm = e^\pm, \mu^\pm$),
and two neutrinos are considered in the analysis.
The number of signal events is expected to be small if two of these three $Z_{(1)}^{(\star)}$ bosons
are reconstructed in leptonic modes. 
Therefore we reconstruct only one of three $Z_{(1)}^{(\star)}$ bosons in the leptonic mode,
and other two in hadronic or neutrino modes. In this paper four channels are studied and the final precision
of the Higgs width measurement is calculated combining accuracies obtained in these channels.

\section{\label{sec:sample}MC samples and analysis tools}

In this analysis we study the following subprocesses:

\begin{eqnarray}\label{eq:channel1}
  e^+e^- & \to Z_1(\lowercase{q_1 q_2})\,H,  \ \ \  H \to Z(\lowercase{q_3 q_4}) Z^\star(\lowercase{\ell_1 \ell_2}) \\
  e^+e^- & \to Z_1(\lowercase{q_1 q_2})\,H,   \ \ \  H \to Z(\lowercase{\ell_1 \ell_2}) Z^\star(\lowercase{q_3 q_4})  \\
  e^+e^- & \to Z_1(\lowercase{\nu \bar{\nu}})\,H, \ \ \    H \to Z(\lowercase{q_1 q_2}) Z^\star(\lowercase{\ell_1 \ell_2}) \\
  e^+e^- & \to Z_1(\lowercase{\nu \bar{\nu}})\,H,  \ \ \  H \to Z(\lowercase{\ell_1 \ell_2}) Z^\star(\lowercase{q_1 q_2})
\end{eqnarray}

The official Monte Carlo (MC) data samples produced by the ILD group are used.
All processes are generated using Whizard 2.8.5 package~\cite{whizard} with the LCIO~\cite{lcio} output format; hadronization
is performed by Pythia6~\cite{pythia6}. The detailed simulation of the ILD detector effects is performed
using the \texttt{ILD\char`_l5\char`_o1\char`_v02} model from the ILCSoft toolkit~\cite{ilcsoft} \texttt{v02-00-02}
using the DD4HEP~\cite{dd4hep} software package. Finally, the events are reconstructed with the MarlinReco~\cite{marlin} package.

The official MC samples are generated assuming four possible combinations with 100$\,\%$ beam polarization, 
${\cal{P}}_{e^-e^+} = (\pm\,1.0, \pm\,1.0)$, and 250 GeV center-of-mass energy.
For the signal events two sets of MC samples are obtained: $e^-_L e^+_R$ (LR) with left-handed electrons
and right-handed positrons and $e^-_R e^+_L$ (RL) with right-handed electrons and left-handed positrons. 
Initial state radiation (ISR) and beam radiation processes are properly included at the generation level. 
The $\gamma\gamma$ beam induced processes are overlaid on the generated events before reconstruction.
The MC samples contain data arrays, so-called data collections, with information about all particles in an event.
In particular, the MCParticles~\cite{mcp} and PandoraPFOs (the Particle Flow Objects reconstructed
with PandoraPFA~\cite{pfo}) collections are included in the samples. 
Table~\ref{tab:taba} shows the basic information taken from logbook ELOG~\cite{elog} for MC samples
selected from repository for this analysis. Zero background contribution is obtained in studies of the MC samples $ZZ(4j)+\gamma^\star$($2\ell$), $ZZ$/$WW(4j)+2\nu$, $H(all)+X$, and $Z(2\nu)Z(2j)+\gamma^\star$($2 \ell$), which are not included in Table~\ref{tab:taba}.

\renewcommand{\arraystretch}{1.2}
\begin{table*}[htb]
\caption{The basic information for all MC samples used in the analysis.
The numbers are taken from generation logbook ELOG. The given cross sections are corrected 
for the decay branching fractions indicated in the first column. The upper and down quarks are labelled as $q_u$ and $q_d$ respectively.}
\label{tab:taba}
\begin{center}
\begin{tabular}
{@{}l@{\hspace{0.1cm}} @{\hspace{0.1cm}}c@{\hspace{0.05cm}} @{\hspace{0.1cm}}c@{\hspace{0.05cm}} @{\hspace{0.1cm}}c@{\hspace{0.05cm}} @{\hspace{0.05cm}}c@{\hspace{0.1cm}} @{\hspace{0.6cm}}c@{\hspace{0.05cm}} @{\hspace{0.1cm}}c@{\hspace{0.05cm}} @{\hspace{0.1cm}}c@{\hspace{0.05cm}} @{\hspace{0.1cm}}c@{\hspace{0.05cm}} @{\hspace{0.6cm}}c@{\hspace{0.05cm}} @{\hspace{0.1cm}}c@{\hspace{0.05cm}} @{\hspace{0.1cm}}c@{\hspace{0.05cm}} @{\hspace{0.1cm}}c@{\hspace{0.01cm}}}
\hline \hline
Process & \multicolumn{4}{c}{Integrated luminosity, $ab^{-1}$} & \multicolumn{4}{c}{Cross section, $fb$} & \multicolumn{4}{c}{Number of events} \\
  & eLpR & eRpL & eLpL & eRpR & eLpR & eRpL & eLpL & eRpR & eLpR & eRpL & eLpL & eRpR \\
\hline
 & \multicolumn{12}{c} {Signal samples}\\
$q\bar{q} H(ZZ) $ & 55.6 & 86.9 & & & 8.99 & 5.75 & & & 5$\cdot$10$^5$ & 5$\cdot$10$^5$ & & \\
$\nu_e \bar{\nu}_e H(ZZ)$ & 316 & 889 & & & 1.58 & 0.56 & & & 5$\cdot$10$^5$ & 5$\cdot$10$^5$ & & \\
$\nu_{\mu \tau} \bar{\nu}_{\mu \tau} H(ZZ)$ & 284 & 445 & & & 1.76 & 1.12 & & & 5$\cdot$10$^5$ & 5$\cdot$10$^5$ & & \\
 & \multicolumn{12}{c} {Background samples}\\
$q\bar{q}$ & 5.00 & 5.00 & & & 128$\cdot$10$^3$ & 70.4$\cdot$10$^3$ & & & 6.40$\cdot$10$^8$ & 3.52$\cdot$10$^8$ & & \\
$W(q\bar{q})W(e+\nu)$ & 5.00 & 5.77 & 1.05 & 1.05 & 10.3$\cdot$10$^3$ & 86.7 & 191 & 191 & 51.4$\cdot$10$^6$ & 5$\cdot$10$^5$ & 2$\cdot$10$^5$ & 2$\cdot$10$^5$ \\
$W(q\bar{q})W(\mu$/${\tau}+\nu)$ & 5.00 & 5.19 & & & 18.8$\cdot$10$^3$ & 173 & & & 93.9$\cdot$10$^6$ & 9$\cdot$10$^5$ & & \\
$Z(q\bar{q})Z(e^+e^-)$ & 5.06 & 5.00 & 1.04 & 1.04 & 1423 & 1219 & 1156 & 1157 & 72$\cdot$10$^5$ & 61$\cdot$10$^5$ & 12$\cdot$10$^5$ & 12$\cdot$10$^5$ \\
$Z(q\bar{q})Z({\mu}^+{\mu}^-$/${\tau}^+{\tau}^-)$ & 5.01 & 5.14 & & & 838 & 467 & & & 42$\cdot$10$^5$ & 24$\cdot$10$^5$ & & \\
$WW$/$ZZ\, (q_u\bar{q}_d\bar{q}_uq_d)$ & 5.00 & 5.32 & & & 12.4$\cdot$10$^3$ & 225 & & & 62$\cdot$10$^6$ & 12$\cdot$10$^5$ & & \\
$WW(other\, 4q)$ & 5.00 & 5.12 & & & 14.8$\cdot$10$^3$ & 225 & & & 74.4$\cdot$10$^6$ & 7$\cdot$10$^5$ & & \\
$ZZ(other\, 4q)$ & 5.05 & 5.11 & & & 1406 & 607 & & & 7.1$\cdot$10$^6$ & 3.1$\cdot$10$^6$ & & \\
$WW(4q)+\gamma^\star$($e^+e^-$) & 28.1 & 168 & 183 & 182 & 0.71 & 0.12 & 0.11 & 0.11 & 2$\cdot$10$^4$ & 2$\cdot$10$^4$ & 2$\cdot$10$^4$ & 2$\cdot$10$^4$ \\
$WW(4q)+\gamma^\star$(${\mu}^+{\mu}^-$/${\tau}^+{\tau}^-$) & 16.9 & 407 & & & 1.19 & 0.05 & & & 2$\cdot$10$^4$ & 2$\cdot$10$^4$ & & \\
$q\bar{q}H(all)$ & 1.5 & 2.3 & & & 343 & 219 & & & 5$\cdot$10$^5$ & 5$\cdot$10$^5$ & & \\
\hline \hline
\end{tabular}
\end{center}
\end{table*}

ILCSoft includes the Marlin software package, which contains special program codes,
so-called processors, used, in particular, for the separation of isolated leptons and the jet reconstruction.
The Marlin package provides an option to use the FastJet~\cite{fj} software codes, designed for clustering particles
in jets based on various reconstruction algorithms. This method is used in the analysis.

\section{\label{sec:selection} Event preselection and initial analysis}

The MC samples studied are preselected using the \mbox{MCParticle} collection containing information
from the MC event generator level. The signal samples are extracted requiring only specific
process and decay chains. The background samples are studied without preselections, however 
the most dangerous background processes are also preselected and studied separately.
All following selections are applied using the information on the reconstruction level.

The first step of the event selection is to identify two isolated lepton candidates.
The standard \texttt{IsolatedLeptonTagging}~\cite{ilt} processor is applied for this goal.
This processor finds high energy leptons in events using multivariate double-cone method and TMVA~\cite{tmva} machine
learning algorithms. We used the default set of parameters and weights included in this processor.
The $Z^\star$ and $Z$ reconstruction efficiencies in the leptonic modes
in the channel with four jets (two jets) are $\sim$67$\,\%$ ($\sim$72$\,\%$) and $\sim$90$\,\%$ ($\sim$91$\,\%$),
respectively.   
Only events with two identified isolated leptons are kept for the following analysis.
These leptons are excluded from the following jet reconstruction procedure.

Energetic ISR photons can be observed inside the detector, this can affect the analysis. A simple ISR identification procedure~\cite{isr} is applied after the lepton identification procedure. If an ISR photon candidate is found, 
it is removed from the PFOs collection and not used in the jet reconstruction.

The next step is the jet reconstruction which is performed using FastJet clustering tools. For this goal we choose the Valencia~\cite{vlc} algorithm, which was specially developed for jet reconstruction at electron-positron colliders.
We select this algroithm for its high efficiency of jet
reconstruction near the beam direction.
After excluding isolated leptons and ISR photons all remaining particles in an event 
are expected to be a part of jets or to come from a $\gamma\gamma$ low $P_t$ process.
On average about 0.4 $\gamma\gamma$ low $P_t$ hadron events are expected per bunch-crossing
at $\,\sqrt{s}$ = 250~GeV~\cite{lowpt}. We use an exclusive $k_T$ clustering
algorithm~\cite{kt} with a generalized jet radius of 1.5 to remove the $\gamma\gamma$ overlay particles
and the Valencia algorithm to force the remaining particles into two or four jets, depending on the studied channel.
Three parameters should be adjusted in the Valencia algorithm, the generalized jet cone radius $R$,
and the $\beta$ and $\gamma$ parameters, which are used to control the clustering order and the background resilience.
We set the $\beta$ parameter to~$1.0$ that corresponds to behavior of the $k_T$ algorithm.
The radius $R$ and the parameter $\gamma$ are tuned to optimize the invariant two-jet mass shape from the $Z \to jj$ decay.
The position and the width of the two-jet mass distribution are evaluated for different $R$ and $\gamma$ values 
using the $M(jj)-M(Z)$, $IQR_{34}$, $RMS_{90}$ and Median parameters proposed in~\cite{vlc}. We chose 
the values of these parameters which provide the best $Z$ boson mass reconstruction quality.
Table~\ref{tab2} shows the sets of the best $R$, $\beta$ and $\gamma$ parameters chosen for
the Valencia jet clustering algorithm in each channel.

\renewcommand{\arraystretch}{1.2}
\begin{table}[htb]
\caption{The best Valencia algorithm parameters choosen for the jet reconstruction in different channels.}
\vspace{0.01cm}
\begin{center}
\label{tab2}
\begin{tabular}
{@{\hspace{0.1cm}}l@{\hspace{0.2cm}} @{\hspace{0.3cm}}c@{\hspace{0.2cm}} @{\hspace{0.3cm}}c@{\hspace{0.2cm}} @{\hspace{0.2cm}}c@{\hspace{0.3cm}} @{\hspace{0.3cm}}c@{\hspace{0.1cm}}}
\hline \hline
\vtop{\hbox{\strut Valencia}\hbox{\strut parameters}\hbox{\strut }}&\vtop{\hbox{\strut Z$_1$($jj$),}\hbox{\strut Z($jj$),}\hbox{\strut Z$^*$($\ell\ell$)}}&\vtop{\hbox{\strut Z$_1$($jj$),}\hbox{\strut Z($\ell\ell$),}\hbox{\strut Z$^*$($jj$)}}&\vtop{\hbox{\strut Z$_1$($\nu \bar{\nu}$),}\hbox{\strut Z($jj$),}\hbox{\strut Z$^*$($\ell\ell$)}}&\vtop{\hbox{\strut Z$_1$($\nu \bar{\nu}$),}\hbox{\strut Z($\ell\ell$),}\hbox{\strut Z$^\star$($jj$)}}\\
\hline
$\beta$ & 1.0 & 1.0 & 1.0 & 1.0\\
$\gamma$ & 0.4 & 0.4 & 0.6 & 0.3\\
$R$ & 1.6 & 0.7 & 1.4 & 1.4\\
\hline \hline
\end{tabular}
\end{center}
\end{table}

The product of the cross section and the branching fraction discussed above can be measured experimentally
using the formula: 

\begin{fleqn}[\parindent]
\begin{equation}\label{eq:sigmaa}
\begin{split}
\sigma(e^+&e^- \to H Z_1) \times Br( H \to Z Z^\star) = \\
& {N_{\rm sig}} / ( \mathcal{L}_{\rm int} \cdot \epsilon \cdot Br(Z_1) \cdot Br(Z) \cdot Br(Z^*) )
\end{split}
\end{equation}
\end{fleqn}

\noindent
where $N_{\rm sig}$ is the number of signal events measured in a specific channel, and $\mathcal{L}_{\rm int}$
is the integrated luminosity of a used data sample. For a studied channel
the selection efficiency is denoted by $\epsilon$ and the relevant decay
branching fractions of the $Z$ bosons decays taken from PDG~\cite{pdg}
are denoted as $Br(Z_1)$, $Br(Z)$, and $Br(Z^*)$.

To get the expected number of signal or background events with ${\cal{P}}_{e^-e^+} = (-0.8, +0.3)$
polarization and the integrated luminosity 2 ab$^{-1}$, we apply a weight factor to each event from the MC samples.
The MC samples are generated assuming 100$\,\%$ polarized beams;
the sample nominal integrated luminosities $\mathcal{L}_{\rm nom}$ are given in Table~\ref{tab:taba}.
The weight factor $W$ is calculated as:

\begin{fleqn}[\parindent]
\begin{equation} \label{eq:wfactor}
W = \left[ \frac{1\pm0.8}{2}\cdot\frac{1\pm0.3}{2} \right] \cdot \frac {\rm 2~ab^{-1}} {\mathcal{L}_{\rm nom} }
\end{equation} 
\end{fleqn}

The numbers of initial MC events, the numbers of events remained after lepton identification, the weight factors, and
the final number of weighted events are given in Table~\ref{tab3}.

\renewcommand{\arraystretch}{1.2}
\begin{table}[htb]
\caption{The numbers of signal events before cuts for different final states obtained from MC samples
with different polarizations before and after lepton tagging and reweighting. The integrated luminosity 2 ab$^{-1}$ and polarization ${\cal{P}}_{e^-e^+} = (-0.8, +0.3)$ is assumed.}
\begin{center}
\label{tab3}
\begin{tabular}
{@{\hspace{0.0cm}}l@{\hspace{0.2cm}}c@{\hspace{0.2cm}}c@{\hspace{0.1cm}} @{\hspace{0.2cm}}c@{\hspace{0.1cm}} @{\hspace{0.2cm}}c@{\hspace{0.1cm}} @{\hspace{0.2cm}}c@{\hspace{0.0cm}}}
\hline \hline
Channels & ${\cal{P}}_{e^-e^+}$ & \vtop{\hbox{\strut MC}\hbox{\strut events}} & \vtop{\hbox{\strut Lepton}\hbox{\strut tagging,}\hbox{\strut events}} & \vtop{\hbox{\strut Weight}\hbox{\strut factors}} & \vtop{\hbox{\strut Weighted}\hbox{\strut number}\hbox{\strut of events}}\\
\hline
\vtop{\hbox{\strut Z$_1$($jj$),}\hbox{\strut Z($jj$),}\hbox{\strut Z$^*$($\ell\ell$)}} & \vtop{\hbox{\strut eLpR}\hbox{\strut eRpL}} & \vtop{\hbox{\strut 23989}\hbox{\strut 23845}} & \vtop{\hbox{\strut 16088}\hbox{\strut 16027}} & \vtop{\hbox{\strut $2.1\cdot10^{-2}$}\hbox{\strut $1.3\cdot10^{-3}$}} & \vtop{\hbox{\strut 338}\hbox{\strut 21}}\\
\vspace{0.1cm}
\vtop{\hbox{\strut Z$_1$($jj$),}\hbox{\strut Z($\ell\ell$),}\hbox{\strut Z$^*$($jj$)}} & \vtop{\hbox{\strut eLpR}\hbox{\strut eRpL}} & \vtop{\hbox{\strut 23261}\hbox{\strut 23132}} & \vtop{\hbox{\strut 20879}\hbox{\strut 20664}} & \vtop{\hbox{\strut $2.1\cdot10^{-2}$}\hbox{\strut $1.3\cdot10^{-3}$}} & \vtop{\hbox{\strut 439}\hbox{\strut 27}}\\
\vspace{0.1cm}
\vtop{\hbox{\strut Z$_1$($\nu_e\bar{\nu}_e$),}\hbox{\strut Z($jj$),}\hbox{\strut Z$^\star$($\ell\ell$)}} & \vtop{\hbox{\strut eLpR}\hbox{\strut eRpL}} & \vtop{\hbox{\strut 24044}\hbox{\strut 23910}} & \vtop{\hbox{\strut 17429}\hbox{\strut 17259}} & \vtop{\hbox{\strut $3.7\cdot10^{-3}$}\hbox{\strut $7.9\cdot10^{-5}$}} & \vtop{\hbox{\strut 65}\hbox{\strut 1.4}}\\
\vspace{0.1cm}
\vtop{\hbox{\strut Z$_1$($\nu_e\bar{\nu}_e$),}\hbox{\strut Z($\ell\ell$),}\hbox{\strut Z$^\star$($jj$)}} & \vtop{\hbox{\strut eLpR}\hbox{\strut eRpL}} & \vtop{\hbox{\strut 23059}\hbox{\strut 23096}} & \vtop{\hbox{\strut 21108}\hbox{\strut 21149}} & \vtop{\hbox{\strut $3.7\cdot10^{-3}$}\hbox{\strut $7.9\cdot10^{-5}$}} & \vtop{\hbox{\strut 79}\hbox{\strut 1.7}}\\
\vspace{0.1cm}
\vtop{\hbox{\strut Z$_1$($\nu_{\mu,\tau}\bar{\nu}_{\mu,\tau}$),}\hbox{\strut Z($jj$),}\hbox{\strut Z$^\star$($\ell\ell$)}} & \vtop{\hbox{\strut eLpR}\hbox{\strut eRpL}} & \vtop{\hbox{\strut 23840}\hbox{\strut 23862}} & \vtop{\hbox{\strut 17103}\hbox{\strut 17168}} & \vtop{\hbox{\strut $4.1\cdot10^{-3}$}\hbox{\strut $1.6\cdot10^{-4}$}} & \vtop{\hbox{\strut 71}\hbox{\strut 2.7}}\\
\vspace{0.1cm}
\vtop{\hbox{\strut Z$_1$($\nu_{\mu,\tau}\bar{\nu}_{\mu,\tau}$),}\hbox{\strut Z($\ell\ell$),}\hbox{\strut Z$^\star$($jj$)}} & \vtop{\hbox{\strut eLpR}\hbox{\strut eRpL}} & \vtop{\hbox{\strut 23189}\hbox{\strut 23225}} & \vtop{\hbox{\strut 21168}\hbox{\strut 21246}} & \vtop{\hbox{\strut $4.1\cdot10^{-3}$}\hbox{\strut $1.6\cdot10^{-4}$}} & \vtop{\hbox{\strut 88}\hbox{\strut 3.3}}\\
\hline \hline
\end{tabular}
\end{center}
\end{table}

The number of Higgs boson signal events is obtained by fitting distributions of the invariant mass $M(jj\ell\ell)$.
However for the channels with $Z \to jj$ and $Z^\star \to \lowercase{\ell\ell}$ decays the following formula gives a
better resolution:

\begin{fleqn}[\parindent]
\begin{equation}\label{eq: Mass Difference}
M_{\Delta} = M(jj\ell\ell) - M(jj) + M(Z_{\rm nom})
\end{equation}
\end{fleqn}

\noindent where $M(Z_{\rm nom}) = 91.2$~GeV. This formula results in a narrower Higgs boson mass peak,
because uncertainties of the jet reconstruction are mostly canceled in the mass difference.

\section{\label{sec:final} Results}

The four channels are studied and the signal statistical uncertainties are evaluated.
To suppress backgrounds various cuts are applied as summarized in Table~\ref{tab4}.
First, the signal and background distributions are obtained with the weighted bin contents and uncertainties.
These distributions are fitted to obtain shape parameters separately for the signal and background.
Then, the signal statistical uncertainties are estimated using the obtained distribution shapes and normalizations.
To reproduce the real data distribution, the weighted signal and background distributions are summed,
the content of each bin is rounded to the integer number and the Poisson uncertainties for the bin contents are assumed.
The binned extended maximum likelihood fit method is applied to the combined distributions with the function
including signal and background terms with the fixed shapes determined in the first step and free normalizations.
Finally, the toy MC method is applied to obtain precise estimates for the signal statistical uncertainties.

\subsection{\label{subsec:channel1} Study of the
 \texorpdfstring{{\boldmath \lowercase{$e^+e^-$}$\to Z_1(\lowercase{j_1 j_2}) \, H(Z Z^\star)$} \newline process 
with {\boldmath $Z \to \lowercase{j_3 j_4}$}, {\boldmath $Z^\star \to \ell^+\ell^-$}}{Lg} }

The final state of the first studied channel includes two leptons and four jets. To form the $Z_1$ and $Z$ bosons
from these four jets we calculate $\chi^2$ for six possible two-jet combinations:

\begin{fleqn}[\parindent]
\begin{myequation}\label{eq:chi sq channel 1}
\begin{multlined}
{\chi}^2 = \frac{(M(Z_1)-M(Z_{\rm nom}))^2}{{{\sigma}{^2}_{M_{Z_1}}}} + \frac{(M(Z)-M(Z_{\rm nom}))^2}{{{\sigma}{^2}_{M_Z}}} \\
+ \frac{(P(Z_1)-\overline{P}(Z_{1}))^2}{{{\sigma}{^2}_{P_{Z_1}}}} + \frac{(P(Z+Z^{\star})-\overline{P}(Z_{1}))^2}{{{\sigma}{^2}_{P_{Z+Z^{\star}}}}}
\end{multlined}
\end{myequation}
\end{fleqn}
\noindent where $\overline{P}(Z_{1}) = 60.0$~GeV/$c$ is the mean $Z_{1}$ momentum in the $e^+e^- \to HZ_1$ process
at the 250~GeV center-of-mass energy. All $\sigma$ parameters are the mean widths of corresponding mass or momentum 
distributions on the reconstruction level. The combination with the minimal $\chi^2$ is selected for the following analysis.

\renewcommand{\arraystretch}{1.2}
\begin{table}[htb]
\caption{The sets of selections used for each studied channel are shown.}
\begin{center}
\label{tab4}
\begin{tabular}
{@{\hspace{0.0cm}}l@{\hspace{0.3cm}} @{\hspace{0.2cm}}c@{\hspace{0.2cm}} @{\hspace{0.2cm}}c@{\hspace{0.2cm}} @{\hspace{0.2cm}}c@{\hspace{0.2cm}} @{\hspace{0.4cm}}c@{\hspace{0.0cm}}}
\hline \hline
Selection &\vtop{\hbox{\strut Z$_1$($jj$),}\hbox{\strut Z($jj$),}\hbox{\strut Z$^*$($\ell\ell$)}}&\vtop{\hbox{\strut Z$_1$($jj$),}\hbox{\strut Z($\ell\ell$),}\hbox{\strut Z$^*$($jj$)}}&\vtop{\hbox{\strut Z$_1$($\nu \bar{\nu}$),}\hbox{\strut Z($jj$),}\hbox{\strut Z$^*$($\ell\ell$)}}&\vtop{\hbox{\strut Z$_1$($\nu \bar{\nu}$),}\hbox{\strut Z($\ell\ell$),}\hbox{\strut Z$^\star$($jj$)}}\\
\hline
$M(\ell\ell)$ & [13, 36] & [70, 95] & [13, 34] & [80, 95] \\
\ \ (GeV/$c^2$) & & & &  \\
$M(Z\to jj)$ & $> 70$ & $< 50$ & [80, 113] & [13, 38] \\
\ \ (GeV/$c^2$) & & & &  \\
$M(Z_1\to jj)$ &  $> 70$ & $> 70$ &  &  \\
\ \ (GeV/$c^2$) & & & &  \\
$E(jjjj\ell\ell)$ & [200, 260] & [200, 260] &  & \\
\ \ (GeV) & & & &  \\
$E(jj\ell\ell)$ &  &  & $< 145$ & $< 145$ \\
\ \ (GeV) & & & &  \\
$P_{\rm max}(\ell)$ & $< 32$ &  & $< 40$ & \\
\ \ (GeV/$c$) & & & &  \\
$P_{\rm min}(\ell)$ & $> 9$ &  & $> 8$ & \\
\ \ (GeV/$c$) & & & &  \\
$P_{\rm max}(j_1)$ &  &  &  & $< 22$ \\
\ \ (GeV/$c$) & & & &  \\
$P_{\rm max}(j_2)$ &  &  &  & $< 42$ \\
\ \ (GeV/$c$) & & & &  \\
$P(jj\ell\ell)$ &  &  & [30, 70] & [40, 70] \\
\ \ (GeV/$c$) & & & &  \\
$|cos\,\theta_{vis}|$ &  &  & $< 0.8$ & $< 0.9$ \\
$\Delta\phi_{ZZ^\star}$  &  &  & $< 120$ & $< 140$ \\
\ \ (degree) & & & &  \\
\hline \hline
\end{tabular}
\end{center}
\end{table}

After jet matching is performed, several cuts are applied.
To remove random backgrounds, the full visible energy in the event is required
to lie in the range $200 < E(jjjj\ell^+\ell^-) < 260$ GeV. After this cut, the dominant backgrounds come from
the $e^+e^- \to W^+W^-\gamma^\star$ and $e^+e^- \to ZZ\gamma^\star$ processes, with the off-shell $\gamma^\star$
decaying to two leptons and the $W$ and $Z$ bosons decaying to two jets.
The distribution of the off-shell photons
falls sharply with increasing mass. The invariant mass of the two leptons rarely exceeds 10~GeV/$c^2$, while the mass
of two leptons in the signal events starts from 10~GeV/$c^2$. 
To obtain the best signal significance, the $M(\ell^+\ell^-) > 13$~GeV/$c^2$ cut is applied to suppress these backgrounds.
A small contribution comes from the four jets $W^+W^-$ and $ZZ$ backgrounds, where $b$ or $c$-quarks
decay semileptonically and the produced leptons split off the corresponding jet.
To suppress these backgrounds,
the minimum lepton momenta is required to be larger than 9 GeV/$c$. 
The maximum lepton momentum is required to be \mbox{$P_{\rm max}(\ell) < 32$ GeV/$c$}. 
We also apply the cut $M(jj) > 70$~GeV/$c^2$, which suppresses the contribution from the $H \to Z^{\star}Z^{\star}$ process.
Figure 1 shows the $M_{\Delta}$ distributions for the signal and background events separately (a) and for the sum
of the signal and background events (b), obtained as described above. 

\begin{figure}[!ht]\label{fig:mxfit channel 1} 
\vspace{-0.7cm}
\centering
\includegraphics[scale=0.4]{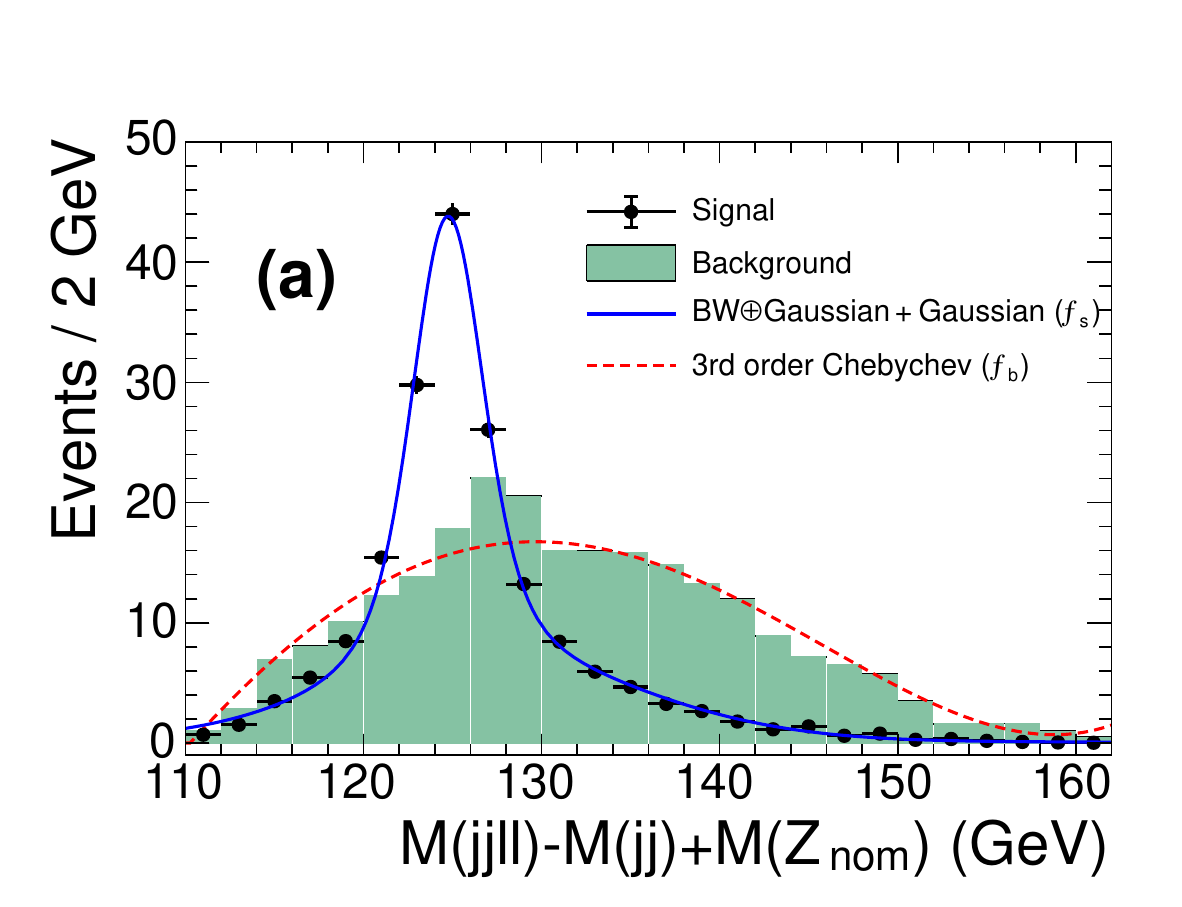}\vspace{-0.2cm}
\includegraphics[scale=0.4]{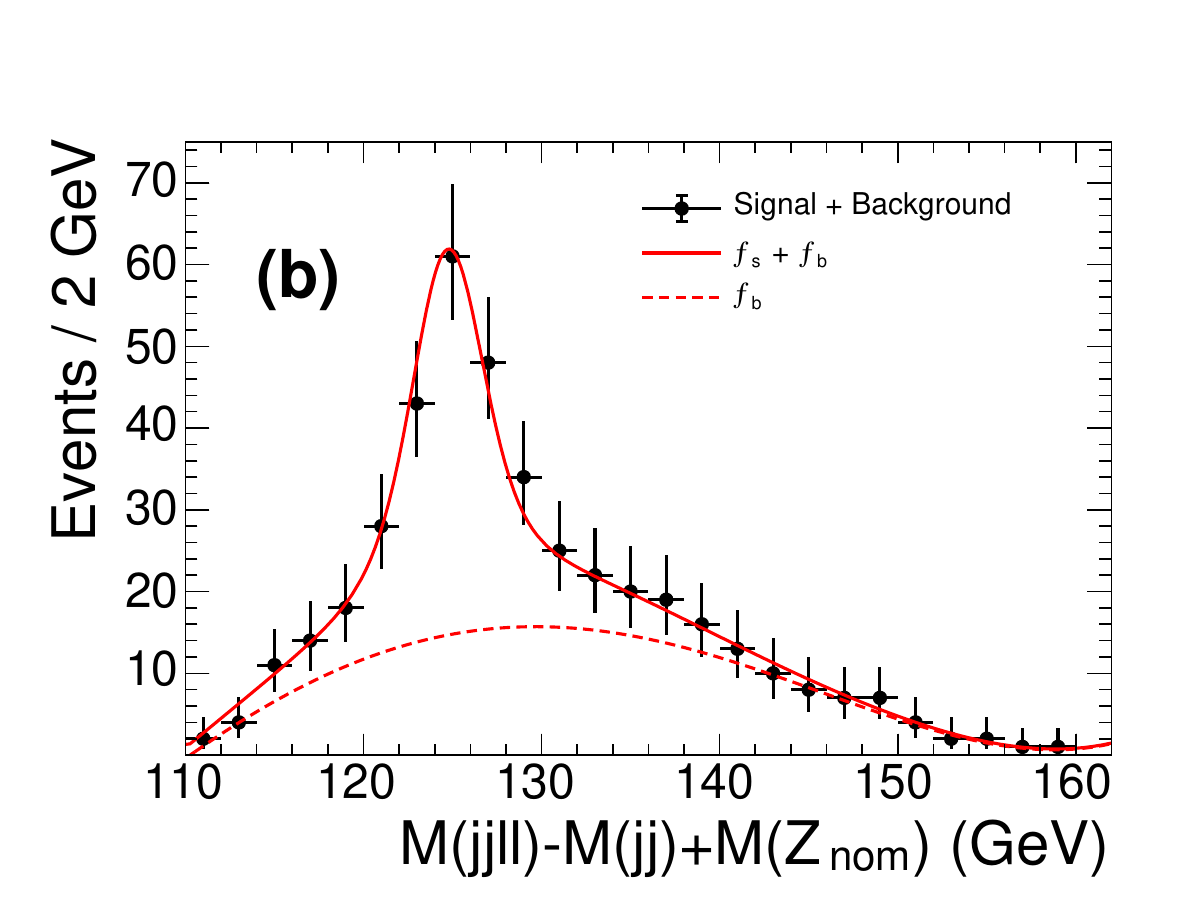}
\caption{The $M_{\Delta} = M(jj\ell\ell) - M(jj) + M(Z_{nom})$ mass distributions are shown for
the $e^+e^- \to Z_1(j_1 j_2) \, H(Z Z^\star)$ process followed by the decays $Z \to j_3 j_4$ and $Z^\star \to \ell^+\ell^-$.
(a)~The distributions are presented separately for the signal (full dots) and background (shaded histogram).
The fit results are overlaid: a blue solid curve for the signal and a red dashed curve for background. 
(b) The sum of the signal and background contributions is shown by full dots together with the
fit results: red dashed curve for background and the red solid curve for the sum.
The functions and the fit methods are described in the text.} 
\end{figure}

The signal distribution is modeled by the sum of two functions: a Breit-Wigner function convolved with a Gaussian function and
an additional wide Gaussian function to account for residual $Z^{\star}Z^{\star}$ events and a few events
due to a wrong jet matching in the ${\chi}^2$ selection. The width of the Breit-Wigner function is 
fixed to the value $\Gamma$ = 2.495~GeV$/c^2$, because the $Z$ boson natural width transfers into the $M_{\Delta}$ value.
The background is described by a third order Chebychev polynomial function.
First, the signal and background distributions are fitted separately to obtain the shapes of the distributions.
Then, the distribution of the sum of the signal and background contributions is fitted with the sum of the corresponding
functions with fixed shapes and free normalizations.
A clear signal peak is observed in the combined distribution.
The fit yields 193.4$\,\pm\,$24.5 signal events.

The two jet mass distributions for the $Z_1$ and $Z$ bosons are shown in Fig.~2.
These distributions are wide, however the large uncertainties in the jet reconstruction mostly cancel in the $M_{\Delta}$ distribution.

\begin{figure}[!ht]\label{fig:MZ channel 1} 
\centering
\includegraphics[scale=0.4]{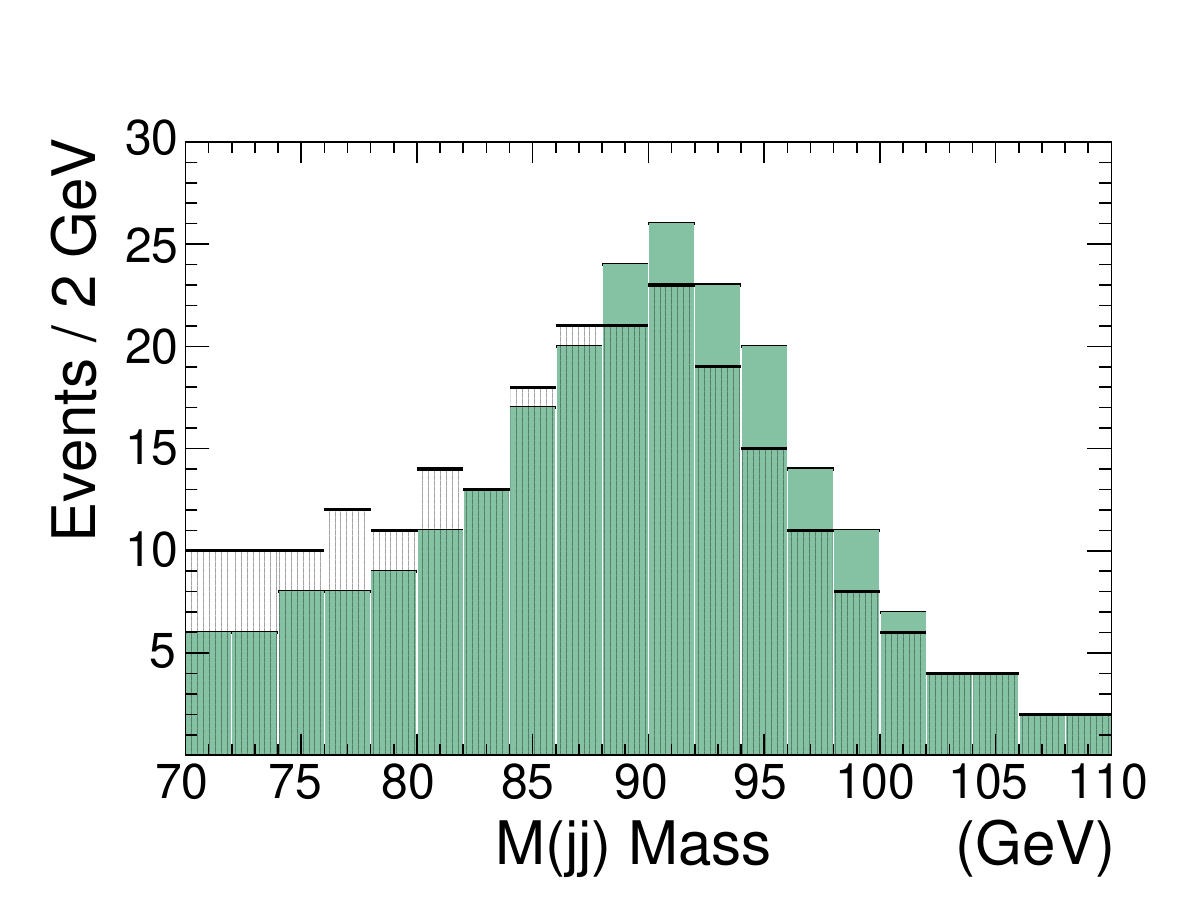}
\caption{The $M(j_1 j_2)$ (hatched histogram) and $M(j_3 j_4)$ (shaded histogram) mass distributions are shown for
the $e^+e^- \to Z_1(j_1 j_2) \, H(Z Z^\star)$ process followed by the decays $Z \to j_3 j_4$ and $Z^\star \to \ell^+\ell^-$.} 
\end{figure}

The signal significance is checked with a toy MC using the RooFit package.
The 10000 $M_{\Delta}$ mass distributions are generated
using the shapes and normalizations for the sum of the signal and background distributions obtained separately.
The generated mass distributions are fitted with a function including both signal and background terms with free normalizations.
Figure 3 demonstrates the distribution of the numbers of the signal events obtained from the toy MC.
The fit of this distribution to the Gaussian function gives the mean value and width of 192.4$\,\pm\,$0.3
and 24.9$\,\pm\,$0.2 events, respectively. 
The toy MC results agree within uncertainties with the combined fit results.
Therefore the statistical uncertainty for this channel is 12.9$\%$.
We quote as the final results the toy MC results for this and other channels.

\begin{figure}[!ht]\label{fig:Toy} 
\centering
\includegraphics[scale=0.4]{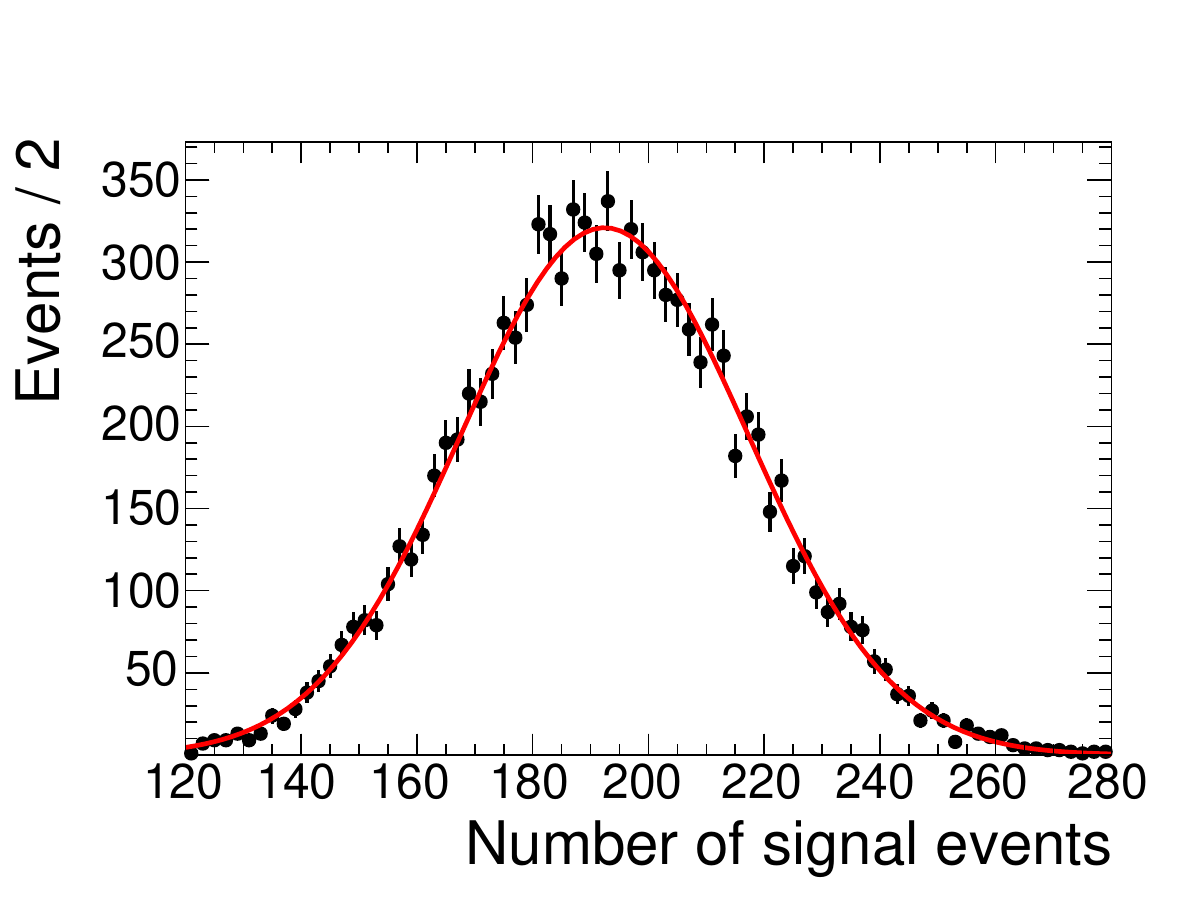}
\caption{The distribution of the number of the signal events obtained from the toy MC fits (dots with errors) is shown
together with a fit by the Gaussian function (curve) as described in the text.}
\end{figure}

\subsection{\label{subsec:channel2} Study of the
\texorpdfstring{{\boldmath \lowercase{$e^+e^-$}$\to Z_1(\lowercase{j_1 j_2}) \, H(Z Z^\star)$} \newline process 
with {\boldmath $Z \to \ell^+\ell^-$}, {\boldmath $Z^\star \to \lowercase{j_3 j_4}$}}{Lg} }

\vspace{-0.2cm}
In this channel the $Z$ boson is reconstructed in the leptonic mode and the $Z^\star$ boson is reconstructed 
in the hadronic mode. The minimal ${\chi}^2$ value is calculated from six possible jet combinations
using the formula:

\begin{fleqn}[\parindent]
\begin{myequation}\label{eq:chi sq channel 2}
\begin{multlined}
{\chi}^2 = \frac{(M(Z_1)-M(Z_{\rm nom}))^2}{{{\sigma}{^2}_{M_{Z_1}}}} + \frac{(E(Z_1)-\overline{E}(Z_{1}))^2}{{{\sigma}{^2}_{E_{Z_1}}}} \\
+ \frac{(P(Z_1)-\overline{P}(Z_{1}))^2}{{{\sigma}{^2}_{P_{Z_1}}}} + \frac{(P(Z+Z^{\star})-\overline{P}(Z_{1}))^2}{{{\sigma}{^2}_{P_{Z+Z^{\star}}}}}
\end{multlined}
\end{myequation}
\end{fleqn}
\noindent where additionally to the $M(Z_{\rm nom})$ and $\overline{P}(Z_{1})$ values defined above,
the mean $Z_{1}$ energy
$\overline{E}(Z_{1}) = 110.0$~GeV is introduced for the $e^+e^- \to HZ_1$ process
at 250~GeV. The energy term slightly improves the quality of the ${\chi}^2$ selection.
All $\sigma$ parameters are obtained using the corresponding distributions
on the reconstuction level.

We apply $70 < M(\ell\ell) < 95$~GeV/$c^2$ and $200 < E(jjjj\ell\ell) < 260$~GeV requirements to
suppress backgrounds due to random lepton pairs and possible $H \to Z^{\star}Z^{\star}$ contribution.
Kinematically, uncorrelated lepton pairs with masses in the $Z$ boson mass region are rarely produced at $\sqrt{s}$ = 250~GeV.
The cut on the di-lepton mass removes almost all combinatorial backgrounds.
Additionally we reject candidates with the mass $M(jj) > 50$~GeV/$c^2$ corresponding to the $Z^\star \to jj$ decay. 
We found no significant remaining backgrounds in this channel after the application of all cuts.
Figure 4 shows the distribution of the mass $M(jj\ell\ell)$, which peaks around of the Higgs boson mass.
The integral of the signal distribution in the mass range $100 < M(jj\ell\ell) < 160$~GeV/$c^2$ is 275.3~events.
The background is very small, the integral over all bins is 18.3 events. 
This background is flat and can be subtracted from the final number of events.
Using this method, the signal mean value and uncertainty are estimated to be 275.3$\,\pm\,$17.2 events.
The statistical uncertainty for this channel is 6.3$\%$.

\begin{figure}[!h]\label{fig:mxfit channel 2} 
\centering
\includegraphics[scale=0.4]{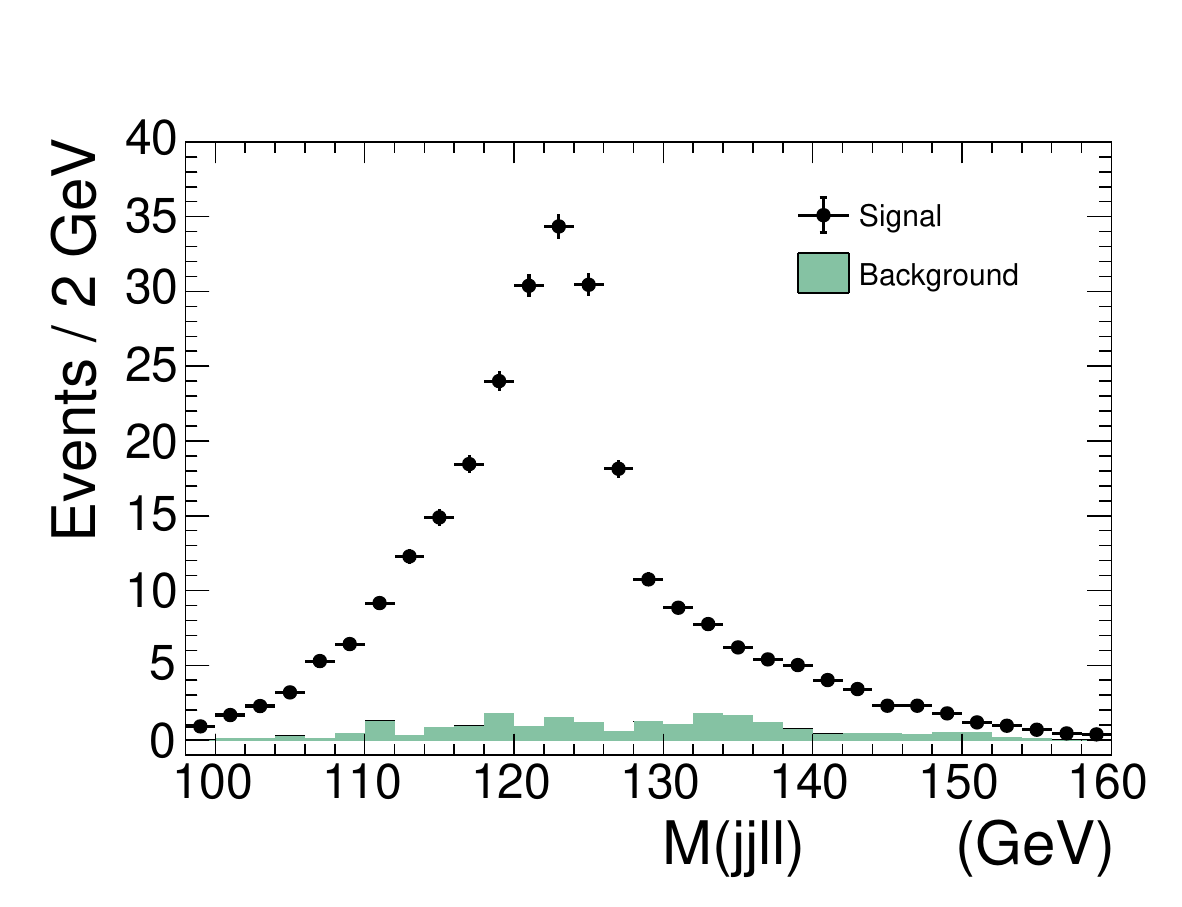}
\caption{The $M(jj\ell\ell)$ mass distributions are shown for the $e^+e^- \to Z_1(j_1 j_2) \, H(Z Z^\star)$ process
followed by the decays $Z \to \ell^+\ell^-$ and $Z^\star \to j_3 j_4$.
The distributions are presented separately for the signal (full dots) and background (shaded histogram).}
\end{figure}

\subsection{\label{subsec:channel3} Study of the
\texorpdfstring{{\boldmath \lowercase{$e^+e^-$}$\to Z_1(\lowercase{\nu \bar{\nu}}) \, H(Z Z^\star)$} \newline process 
with {\boldmath $Z \to \lowercase{jj}$}, {\boldmath $Z^\star \to \ell^+\ell^-$}}{Lg} }

We also studied the processes in which the directly produced $Z_1$ boson decays to neutrinos. 
Comparing with decays in the hadronic mode, a smaller number of signal events is expected in the neutrino mode,
due to the smaller $Z$ decay branching fraction. However the final state has a simple signature with only 
two jets and two leptons. 

We studied different background sources to this channel with the $Z\to jj$ and $Z^\star\to\ell^+\ell^-$ decays.
There are many background sources with large cross sections which can contribute to this channel. 
Special attention must be paid to the $e^+e^- \to Z(2j) Z(\tau^+\tau^-)$ process with  
leptonic $\tau$ decays, and also to the $e^+e^- \to W(2j) W(\ell \nu)$ process with a lepton produced within one of the jets. Another potentially dangerous background is due to $b \bar{b}$ pair production, where both $b$-quarks decay semileptonically.
These two leptons have to split off the jets to imitate the studied configuration. The probability for two leptons produced in hadronic jets to be identified
as isolated leptons is very low, however it is compensated by high rates for this process.
These backgrounds have a signature similar to the signal configuration. To reduce these backgrounds a set of cuts
given in Table~\ref{tab4} is applied.
The effective cuts are on the full visble momentum $30 < P(jj\ell\ell) < 70$~GeV$/c$ and energy $E(jj\ell\ell) < 145$~GeV$/c$.
The angular cuts on the azimuthal angle of the full system relative to the beam direction, $|cos\,\theta_{vis}| < 0.8$,
and on the angle between the $Z$ and $Z^\star$
boson directions, $\Delta\phi_{ZZ^\star} < 120^\circ$, are used to further suppress the backgrounds.
Some additional suppression of specific background configurations can be achieved if dedicated cuts are applied on the
minimum and maximum momenta of the leptons.  
We also tested the processes $e^+e^- \to b\bar{b}$ and $e^+e^- \to Z H(b\bar{b})$ and found a small background contribution.
To suppress these backgrounds we applied the $13 < M(\ell\ell) < 34$~GeV$/c^2$ cut.
The cut $80 < M(jj) < 113$~GeV$/c^2$ is used to suppress the contribution from the $H \to Z^{\star}Z^{\star}$
process. 
Figure~5 shows the $M_{\Delta}$ distributions for the signal and background events separately (a) and their sum (b).

\begin{figure}[!ht]\label{fig:mxfit channel 3} 
\vspace{-0.7cm}
\centering
\includegraphics[scale=0.4]{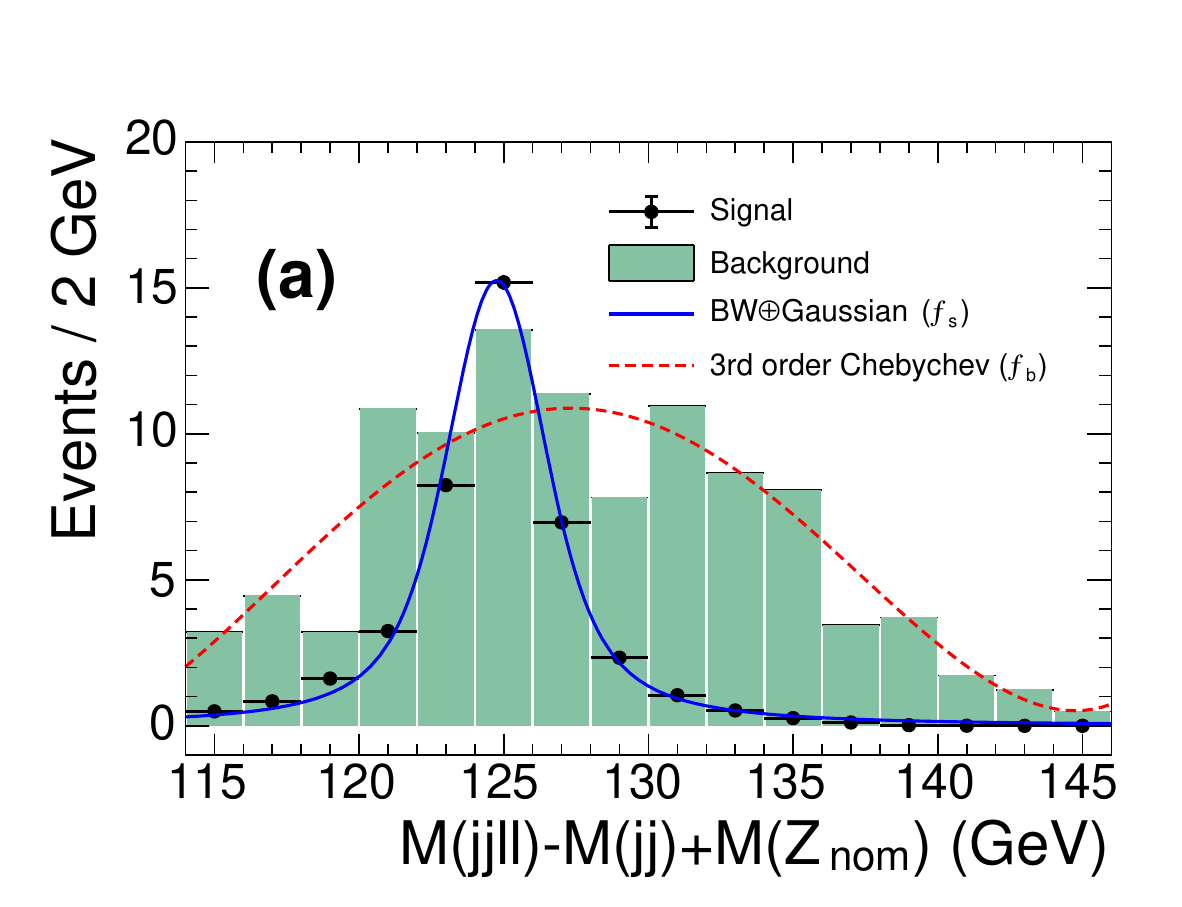}\vspace{-0.1cm}
\includegraphics[scale=0.4]{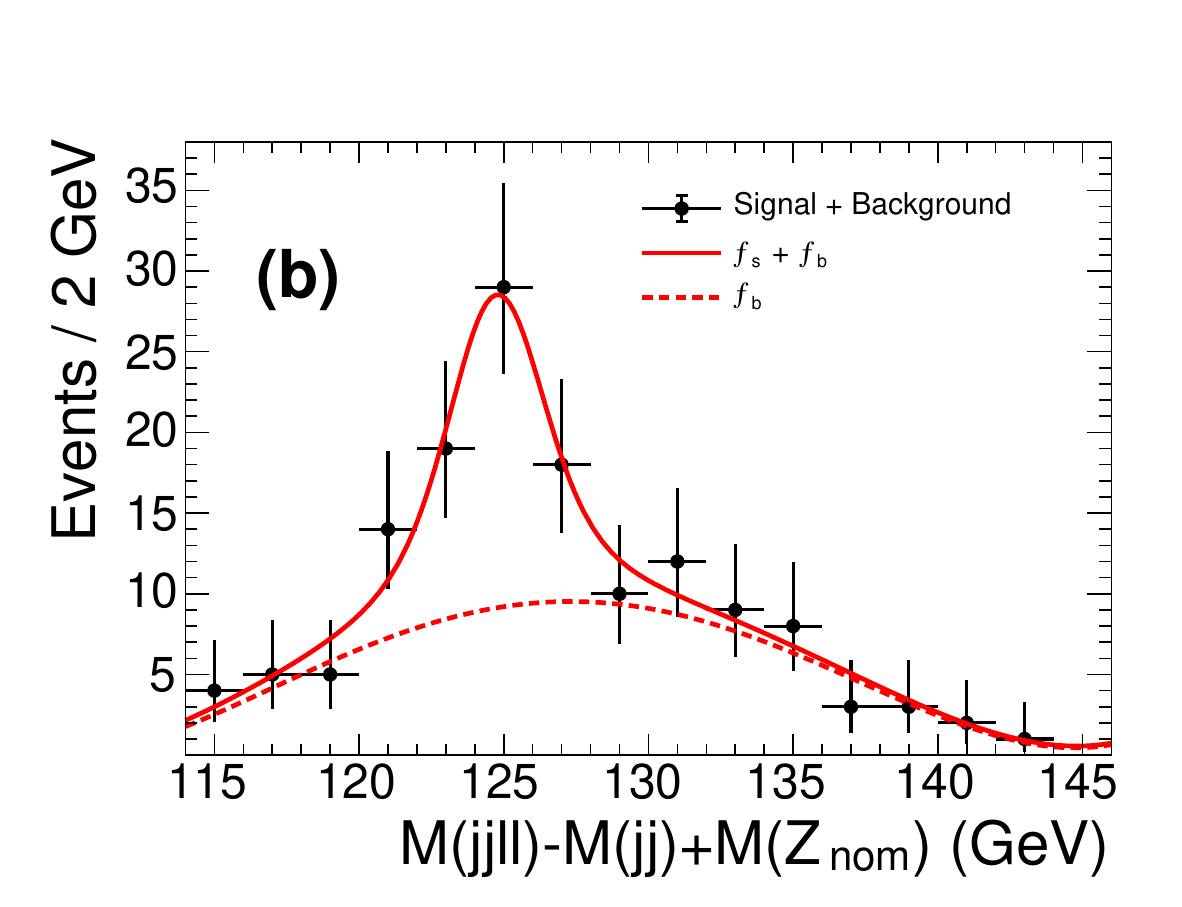}
\caption{The $M_{\Delta} = M(jj\ell\ell) - M(jj) + M(Z_{nom})$ mass distributions are shown for
the $e^+e^- \to Z_1(\nu \bar{\nu}) \, H(Z Z^\star)$ process followed by the decays $Z \to jj$
and $Z^\star \to \ell^+\ell^-$.
(a)~The distributions are presented separately for the signal (full dots) and background (shaded histogram).
The fit results are overlaid: a blue solid curve for the signal and a red dashed curve for background. 
(b) The sum of the signal and background contributions is shown by full dots together with the
fit results: red dashed curve for background and the red solid curve for the sum.
The functions and the fit methods are described in the text.}
\end{figure}

The fit procedure is applied to estimate the statistical significance.
The signal is modeled with a convolution of a Breit-Wigner function and a Gaussian function.
The width of the Breit-Wigner function is fixed to the value $\Gamma$ = 2.495~GeV$/c^2$.
The background is described by the third order Chebychev polynomial function. First, the signal and background distributions are fitted
separately to obtain the shapes of the functions. Then the sum of the signal and background contributions
is fitted by the sum of corresponding functions with fixed shapes and free normalizations.
The fit results are shown in Fig.~5. Finally, the fit gives 52.0$\,\pm\,$12.7 signal events. 
The toy MC estimation gives 51.9$\,\pm\,$13.0 events, that results in the statistical uncertainty of 25.1$\,\%$.

\subsection{\label{subsec:channel4} Study of the
\texorpdfstring{{\boldmath \lowercase{$e^+e^-$}$\to Z_1(\lowercase{\nu \bar{\nu}}) \, H(Z Z^\star)$} \newline process 
with {\boldmath $Z \to \ell^+\ell^-$}, {\boldmath $Z^\star \to \lowercase{jj}$}}{Lg} }

As in the previous channel, the $Z_1$ boson here decays also to neutrinos. However the hadronic and leptonic
modes for the $Z$ and $Z^\star$ bosons are swapped.

The dangerous background sources are similar to the previous channel, except the $b \bar{b}$ background.
However this channel's backgrounds are more effectively suppressed due to the large dilepton mass.
We select events in the intervals $80 < M(\ell\ell) < 95$~GeV$/c^2$, $13 < M(jj) < 38$~GeV$/c^2$,
$40 < P(jj\ell\ell) < 70$~GeV$/c$ and $E(jj\ell\ell) < 145$~GeV to suppress random lepton pairs and 
other backgrounds. 
Angular cuts $|cos\,\theta_{vis}| < 0.9$ and $\Delta\phi_{ZZ^\star} < 140^\circ$ are also applied.
Finally we require that at least one jet from the $Z^\star$ decays has the momentum $< 22$~GeV$/c$,
whereas the second one has the momentum $< 42$~GeV$/c$.
Figure~6 shows the mass distribution $M(jj\ell\ell)$ obtained after all the cuts applied for the signal and background events
separately (a) and their sum (b).

\begin{figure}[!ht]\label{fig:mxfit channel 4} 
\centering
\includegraphics[scale=0.4]{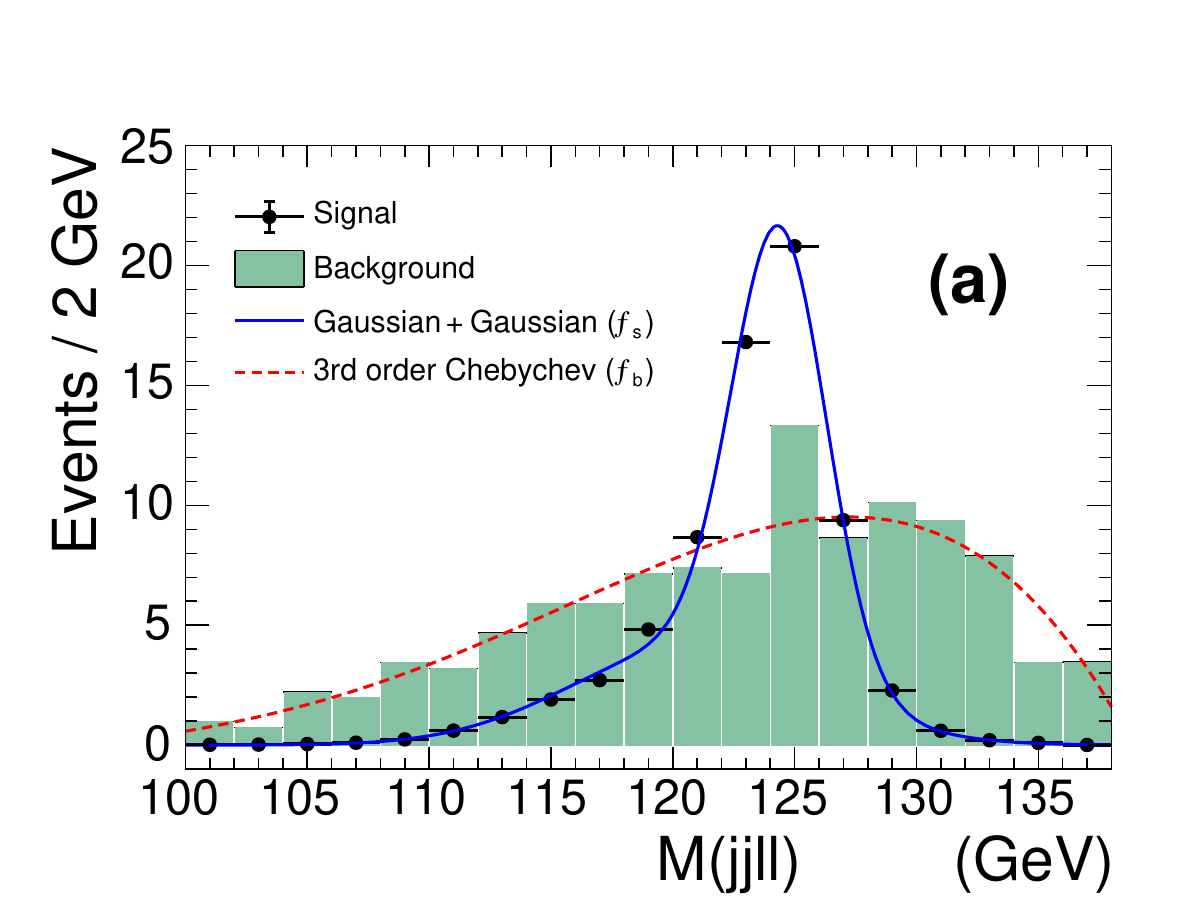}\vspace{-0.1cm}
\includegraphics[scale=0.4]{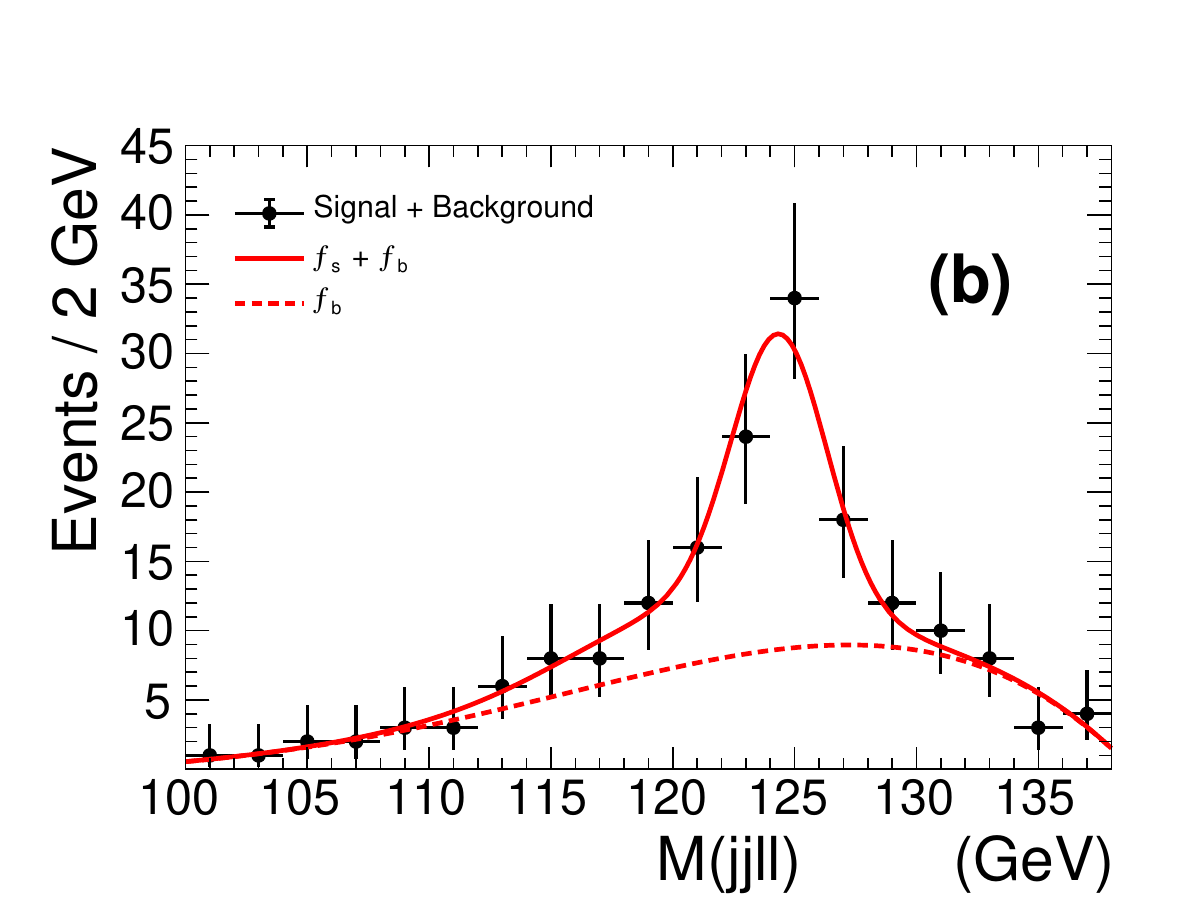}
\caption{The $M(jj\ell\ell)$ mass distribution is shown for the $e^+e^- \to Z_1(\nu \bar{\nu}) \, H(Z Z^\star)$ process
followed by the decays $Z \to \ell^+\ell^-$ and $Z^\star \to jj$.
(a)~The distributions are presented separately for the signal (full dots) and background (shaded histogram).
The fit results are overlaid: a blue solid curve for the signal and a red dashed curve for background. 
(b) The sum of the signal and background contributions is shown by full dots together with the
fit results: red dashed curve for background and the red solid curve for the sum.
The functions and the fit methods are described in the text.}
\end{figure}

\renewcommand{\arraystretch}{1.2}
\begin{table*}[ht]
\vspace{0.3cm}
\caption{The fitted number of signal events and their relative statistical uncertainties obtained from the toy MC for each channel.
The relative statistical uncertainties for the fitted number of signal events correspond directly to the relative statistical
uncertainties for $\sigma(e^+e^- \to HZ) \times {\cal B}r(H \to ZZ^*)$.}
\begin{center}
\label{tab5}
\begin{tabular}
{@{\hspace{0.0cm}}l@{\hspace{0.3cm}}c@{\hspace{0.6cm}}c@{\hspace{0.5cm}}@{\hspace{0.1cm}}c@{\hspace{0.5cm}} @{\hspace{0.1cm}}c@{\hspace{0.4cm}} @{\hspace{0.1cm}} @{\hspace{0.0cm}}c}
\hline \hline
 & \vtop{\hbox{\strut Z$_1$($jj$), Z($jj$),} \hbox{\hspace{0.5cm}\strut Z$^*$($\ell\ell$)}} & \vtop{\hbox{\strut Z$_1$($jj$), Z($\ell\ell$),}\hbox{\hspace{0.5cm}\strut Z$^*$($jj$)}} & \vtop{\hbox{\strut Z$_1$($\nu \bar{\nu}$), Z($jj$),} \hbox{\hspace{0.5cm}\strut Z$^*$($\ell\ell$)}} & \vtop{\hbox{\strut Z$_1$($\nu \bar{\nu}$), Z($\ell\ell$),} \hbox{\hspace{0.5cm}\strut Z$^\star$($jj$)}} & Sum\\
\hline
\multicolumn{6}{c} {\vtop{\vspace{-0.2cm}\hbox{\strut 2 ab$^{-1}$ eLpR}\hbox{\strut \vspace{-0.3cm} }}}\\
Number of events & 192.4 $\pm$\,24.9 & 275.3 $\pm$\,17.2 & 51.9 $\pm$\,13.0 & 73.3 $\pm$\,14.2 & - \\
\vtop{\hbox{\strut Statistical} \hbox{\strut uncertainty}} & 12.9\% & 6.3\% & 25.1\% & 19.3\% & 5.29\%\\
\multicolumn{6}{c} {\vtop{\vspace{-0.2cm}\hbox{\strut 0.9 ab$^{-1}$ eLpR + 0.9 ab$^{-1}$ eRpL}\hbox{\strut \vspace{-0.3cm} }}}\\
Number of events & 135.2 $\pm$\,20.4 & 202.2 $\pm$\,14.7 & 30.9 $\pm$\,10.7 & 67.3 $\pm$\,14.3 & - \\
\vtop{\hbox{\strut Statistical} \hbox{\strut uncertainty}} & 15.1\% & 7.3\% & 34.6\% & 21.2\% & 6.15\%\\
\hline \hline
\end{tabular}
\end{center}
\end{table*}

It has to be noted that the mass distribution is relatively narrow in this channel. 
This is because of only two jets and two leptons in the final state. Therefore we do not need to apply the 
minimum $\chi^2$ method and the invariant mass of the two jets will not change if some of particles are wrongly assigned to jets. The signal distribution is described by the sum of two Gaussians.
The wide Gaussian accounts for the $H \to Z^{\star}Z^{\star}$ contribution. The background is described
by the third order Chebychev polynomial function. Using the fit procedure
we obtain 74.1$\,\pm\,$13.9 events.
The toy MC estimation gives 73.3$\,\pm\,$14.2 events,
corresponding to a statistical uncertainty of 19.3$\%$. The combined fit results perfectly agree with the
toy MC values in all four channels.

\section{\label{sec:significance} Combined signal significance estimate}

An important result of this study is an estimate of accuracy which can be reached for the Higgs width measurement.
To estimate the accuracy, we calculate the combined statistical uncertainty
for the four studied channels using the formula $S_{\rm comb} = 1 / \sqrt{\sum_{i=1}^4 S_i^{-2}}$.
Results obtained for all studied channels
and the combined value of statistical uncertainty are given in Table~\ref{tab5}.

As given in Table~\ref{tab5}, the statistical uncertainty of the proposed method is 5.29$\,\%$ for
the integrated luminosity 2 ab$^{-1}$ and polarization ${\cal{P}}_{e^-e^+} = (-0.8, +0.3)$.
Alternatively, we assumed two data samples with the polarizations ${\cal{P}}_{e^-e^+} = (-0.8, +0.3)$
and ${\cal{P}}_{e^-e^+} = (+0.8, -0.3)$ and the integrated luminosity of 0.9~ab$^{-1}$ each.
The same analysis is repeated for this data taking scheme and the total statistical uncertainty of 6.15$\,\%$ is obtained.

We note that the branching fraction $Br( H \to Z Z^\star)$ will have a small contribution
from the $H \to Z^\star Z^\star$ decay, which is around (5-10)$\,\%$ depending on the studied channels. 
This contribution can be accurately evaluated and corrected for. 
In the last two channels there is also a contribution from to the $W$-fusion
process $e^+e^- \to H(ZZ^\star)\,\nu_e \bar{\nu_e}$. 
For the used cuts its fraction is about 15\,$\%$ of the selected signal events.
As in the case of the previous correction, this fraction can be precisely evaluated and, therefore, does not result in
a loss of accuracy. 

The systematic uncertainties are not studied in this analysis. The largest
systematic uncertainties are expected from the uncertainty in the selection efficiency
and the uncertainty due to the signal and background shape modeling in the fit. The later systematic 
uncertainty will dominate.
Unfortunately accurate estimates of the systematic uncertainties cannot be performed without real data.

\section{\label{sec:conclusions} Conclusions}

We studied the $e^+e^- \to HZ$ process with subsequent $H \to Z Z^\star$ decay. The analysis is performed assuming
the integrated luminosity 2 ab$^{-1}$ collected at the $e^+e^-$ collisions with center-of-mass energy 250 GeV and
the beam polarizations ${\cal{P}}_{e^-e^+} = (-0.8, +0.3)$.
Four channels are studied and the corresponding signal and background contributions are estimated using MC simulation.
Summing results obtained in the four studied channels we obtain the combined statistical uncertainty 5.29$\,\%$. 
This indicates, that the Higgs width can be measured using this method with an accuracy of about (5--6)$\,\%$
within the model-independent approach. We also repeated the analysis assuming two data samples with integrated luminosities 0.9 ab$^{-1}$ and two beam polarizations ${\cal{P}}_{e^-e^+} = (\mp0.8, \pm0.3)$ and obtained the statistical uncertainty of 6.15$\,\%$. The accuracy of this method is similar to one obtained in ~\cite{wp,hww},
where measurements of four or five processes have to be performed to determine the Higgs width.
The results of both methods can be combined to further improve the accuracy.

The Higgs width can be potentially constrained in the future with an accuracy of about 2$\,\%$ by applying a global fit with many Higgs
related parameters included in the frame of the effective field theory (EFT) approach~\cite{eft}. Our measurement can be used to
test the Higgs width value obtained within the SM, as well as within the EFT approach. 
Moreover, the results obtained in this analysis can be included in the global EFT fit, that can improve its accuracy. Quantitatively it will be studied in a global fit for the upcoming Snowmass 2021 Higgs white paper with the input measurement from this paper.

\section*{\label{sec:acknowledgements}ACKNOWLEDGMENTS}
Authors are grateful to I. Bozovic-Jelisavcic, D. Jeans, J. Tian and A. F. Zarnecki for useful discussions.
We would like to thank the LCC generator working group and the ILD software working group for providing the simulation and reconstruction tools and producing the Monte Carlo samples used in this study.
This work has benefited from computing services provided by the ILC Virtual Organization, supported by the national resource providers of the EGI Federation and the Open Science GRID. The work is supported by the Ministry of Science and Higher Education of the Russian Federation, Agreement No. 14.W03.31.0026.

\end{document}